\documentclass{elsart41}
\usepackage{graphicx}
\usepackage{amssymb}

\begin{document}

\begin{frontmatter}



\title{Phase diagram of the excitonic insulator}


\author[label1]{B. H\"ulsen},
\author[label2]{F. X. Bronold\corauthref{Name1}},
\ead{franz.bronold@physik.uni-greifswald.de}
\author[label2]{H. Fehske},
\author[label3]{K. Yonemitsu}

\address[label1]{Fritz-Haber-Institut, D-14195 Berlin, Germany}
\address[label2]{Institut f\"ur Physik, Ernst-Moritz-Arndt-Universit\"at Greifswald, D-17487 Greifswald, Germany}
\address[label3]{Institute for Molecular Science, Myodaiji, Okazaki 444-8585, Japan}

\corauth[Name1]{Corresponding author. Tel: +49 3834-864788, Fax: -864791}

\begin{abstract}
Motivated by recent experiments, which give strong evidence for
an excitonic insulating phase in $\rm TmSe_{0.45}Te_{0.55}$, we developed a 
scheme to quantitatively construct, for generic two-band models, the phase 
diagram of an excitonic insulator. As a first application of our approach,
we calculated the phase diagram for an effective mass two-band model with 
long-range Coulomb 
interaction. The shielded potential approximation is used to derive a generalized 
gap equation controlling for positive (negative) energy gaps the transition from a 
semi-conducting (semi-metallic) phase to an insulating phase. Numerical results,
obtained within the quasi-static approximation, show a steeple-like phase diagram
in contrast to long-standing expectations. 
\end{abstract}

\end{frontmatter}

The possibility of an excitonic insulator (EI) phase, 
separating, below a critical temperature, a semiconducting from a
semi-metallic phase, has been 
predicted by theorists more than three decades ago~\cite{Hal68}.
However, experimental efforts to establish this phase in  
actual materials largely failed. It is only until 
recently, that detailed experimental investigations of 
$\rm TmSe_{0.45}Te_{0.55}$ suggested the existence 
of an EI phase in this compound~\cite{Neu90,Wac04}.
The pressure dependence of the electrical resistivity below $270{\rm K}$,
for instance, strongly points towards an emerging EI phase~\cite{Neu90}. 
Further evidence for collective behavior which may have its origin in an 
EI phase comes from the linear increase of the thermal conductance and 
diffusivity at very low temperatures~\cite{Wac04}.

Under the assumption that the external pressure controls the energy 
gap ${\rm E_g}$, the resistivity data have been used to construct a phase 
diagram for $\rm TmSe_{0.45}Te_{0.55}$ in the ${\rm E_g}$-T plane~\cite{Neu90}.
Although experimental data strongly suggest that this phase diagram is 
the phase diagram of an EI, to unambiguously decide if this interpretation
is correct requires further theoretical examination, taking the relevant parts
of the electronic structure of the material into account. 
However, even for the simplest two-band models, a quantitative phase diagram for 
an EI has never been calculated. As a 
first step towards a theoretical scrutiny of the phases of the $\rm Tm[Se,Te]$ 
system it is therefore appropriate to present here such a calculation.
 
In close analogy to the strong-coupling theory of superconductivity, we 
employed a matrix propagator formalism.  
Within a two-band model, the anomalous or off-diagonal 
(in the band indices $i=1,2$) self-energy $\Sigma_{12}({\bf k},i\omega_n)$
describing the pairing between conduction and valence band electrons 
serves as an order parameter:
$\Sigma_{12}({\bf k},i\omega_n) \neq 0$ signals the existence of the EI phase. 
Our selfconsistent approximation enables us to take a variety of physical processes 
into account and results in a nonlinear functional equation for 
$\Sigma_{12}({\bf k},i\omega_n)$. 
Linearizing this equation in the vicinity of the phase boundary, where 
$\Sigma_{12}({\bf k},i\omega_n)$ is small, yields a generalized ``gap equation''.
The phase boundary ${\rm T_c}({\rm E}_g)$ can then be found by mapping out the 
T-${\rm E}_g$ range for which the ``gap equation'' has nontrivial solutions.

We applied this scheme to an isotropic, effective mass two-band model for  
valence and conduction 
band electrons interacting via the long-range Coulomb potential
$V_0({\bf q})=4\pi e^2/\varepsilon_0q^2$. The energy gap ${\rm E_g}$ 
is indirect ($\Gamma$-X) and can be positive or negative. Within 
the shielded potential approximation the generalized gap equation for 
the real part of the interband self-energy reads 
\begin{eqnarray}
\Delta({\bf k},\tilde{e}_i)\!&=&\!\!\int\!\!\frac{d{\bf k}'}{(2\pi)^3}
\big[
V({\bf k}-\!{\bf k}',\tilde{e}_i,\tilde{e}_1')
B_2(\tilde{e}_2',\tilde{e}_1')\Delta({\bf k}',\tilde{e}_1')\nonumber\\
\!&+&\! V({\bf k}-\!{\bf k}',\tilde{e}_i,\tilde{e}_2')
B_1(\tilde{e}_1',\tilde{e}_2')\Delta({\bf k}',\tilde{e}_2')\big],
\label{full}
\end{eqnarray}
where $\tilde{e}_i({\bf k})=e_i({\bf k})-\mu_i$,
$B_i(x,y)=(x-y)/[(x-y)^2+\gamma_i(\bf{k})^2]$,
and $V({\bf k}-\!{\bf k}',\tilde{e}_i,\tilde{e}_j')=
{\rm Re}V_s^r({\bf k}-\!{\bf k}',\tilde{e}_i-\tilde{e}_j')n_F(\tilde{e}_j')
-{\rm P}\int (d\omega/\pi) 1/(\tilde{e}_i-\tilde{e}_j'-\omega)
{\rm Im}V_s^r({\bf k}-\!{\bf k}',\tilde{e}_i-\tilde {e}_j')n_B(-\omega)$ with 
$V_s^r({\bf q})=V_0({\bf q})/\varepsilon^r({\bf q},\omega)$ the dynamically screened 
Coulomb potential. 
To derive Eq. (\ref{full}) we employed a quasi-particle approximation for the 
intraband propagators with renormalized band dispersions and lifetimes given by
$e_i({\bf k})=\epsilon_i({\bf k})+{\rm Re}\Sigma^r_{ii}({\bf k},e_i({\bf k})-\mu_i)$
and $\gamma_i({\bf k})=-{\rm Im}\Sigma^r_{ii}({\bf k},e_i({\bf k})-\mu_i)$, 
respectively. The chemical potentials $\mu_i$ are measured from the respective band 
extrema.

The full analysis of Eq. (\ref{full}) is the subject of a forthcoming publication~\cite{first}.
Here, we focus on the quasi-static approximation,
$V_s^r({\bf k}-\!{\bf k}',\tilde{e}_i-\tilde{e}_j')\approx V_s^r({\bf k}-\!{\bf k}',0)$, 
which simplifies the gap equation enormously. Except for very
small band overlaps (very small Fermi surfaces), we expect this approximation 
to work reasonably well, as it does for intraband self-energies. For equal 
band masses and temperature independent screening, the quasi-static approximation reduces Eq. (\ref{full}) to
\begin{eqnarray}
\Delta(u)&=&\!\!\int_{u_1}^{\infty} du'V(u,u')\frac{\tanh{u'}}{2u'}\Delta(u')
\label{qsa}\\
V(u,u')&=&\!\!\sqrt{\frac{1}{4\pi^2{\rm k_B T}}}\frac{1}{{\rm k}}
\log\big[\frac{({\rm k}+{\rm k'})^2+\kappa^2}{({\rm k}-{\rm k'})^2+\kappa^2}\big]
\end{eqnarray}
with ${\rm k}=\sqrt{u-u_1}$, ${\rm k}'=\sqrt{u'-u_1}$, $u_1={\rm E_g}/4{\rm k_BT}$, and 
$\kappa^2=(2\sqrt{|{\rm E_g}|}/\pi{\rm k_BT})\theta(-{\rm E_g})$.  
To construct the phase boundary ${\rm T_c}({\rm E}_g$), we discretize Eq. (\ref{qsa})  
and determine, for fixed ${\rm E_g}$, the temperature ${\rm T=T_c}$ for which the determinant of the 
coefficient matrix of the resulting system of linear equations vanishes. For ${\rm E_g}<0$
this approach can be directly applied, whereas for ${\rm E_g}>0$, the logarithmic 
singularity of the kernel has to be removed first~\cite{first}.

\begin{figure}
\centerline{\includegraphics*[width=0.9\columnwidth]{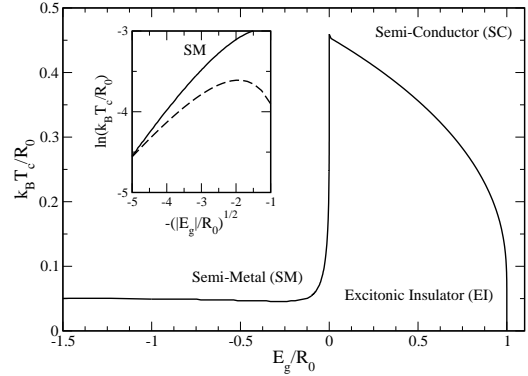}}
\caption{Phase diagram for an excitonic insulator with equal band masses.}
\label{Fig1}
\end{figure}

The phase boundary ${\rm T_c}({\rm E}_g$) is presented in Fig. \ref{Fig1}, measuring energy and 
temperatures in units of the exciton Rydberg $R_0$. Above $T_1\approx 0.45$, 
the EI phase is unstable. Below $T_1$, we find a steeple-like phase boundary which
strongly discriminates between ${\rm E_g}>0$ and  ${\rm E_g}<0$. 
For ${\rm E_g}>0$, $T_c({\rm E}_g$) smoothly
decreases to zero at ${\rm E_g}=1$, the critical band gap, above which the EI phase cannot exist.
For ${\rm E_g}<0$, in contrast, $T_c({\rm E}_g)$ initially drops extremely fast, within a few 
percent of $R_0$, to a second critical temperature $T_2\approx 0.04$. Below $T_2$ the EI phase is 
stable with an almost constant ${\rm T_c(|E_g|)}$, which, however, for larger band overlaps 
slowly decreases (see inset). The  
steeple-like shape of the phase diagram reflects the different phases from 
which the EI is approached: semi-conducting for ${\rm E_g}>0$ and semi-metallic 
for ${\rm E_g}<0$. 
Entering the EI phase 
from the semi-conductor side leads to formation of {\it strongly bound} excitons. 
On the other hand, when the EI phase is approached from the semi-metal, 
exciton formation is strongly suppressed due to the free carrier's screening of 
the Coulomb potential. 
In that case, {\it loosly bound} Cooper-type pairs emerge, resulting in a rather fragile 
EI phase. The crossover from excitons to Cooper-type pairs occurs at ${\rm E_g}\approx -0.3$,
the band overlap, where the screening length becomes roughly equal to the exciton radius.  
For ${\rm |E_g|}\ll 4{\rm k_BT}$, the critical temperature is exponentially small
and approximately given by
${\rm k_BT_c}\approx
(\gamma|{\rm E_g|}/\pi)\exp{(-\pi\sqrt{|{\rm E_g}|}/{\rm ln}{(1+\pi\sqrt{|{\rm E_g}|}/2))}}$, with
$\gamma=\exp{(0.577)}$ (dashed line in the inset). 
Anisotropies in the band structure and other pair breaking effects would 
easily destroy this part of the phase diagram.

Support from SFB 652 is greatly acknowledged.

\end{document}